\documentclass[twocolumn,aps,prl,10pt]{revtex4-2}

\usepackage{amsmath}
\usepackage{amssymb}
\usepackage{wasysym}
\usepackage{graphicx}
\usepackage{hyperref}
\usepackage{newtxtext}
\usepackage[varvw]{newtxmath}
\usepackage{mathtools}
\usepackage[shortlabels]{enumitem}
\usepackage{placeins} 
\usepackage{bbold}
\usepackage{physics}

\hypersetup{
    colorlinks,
    linkcolor={blue},
    citecolor={blue},
    urlcolor={blue}
}

\graphicspath{{./}{./figs/}}

\newcommand{\rme}{\mathrm{e}}
\newcommand{\rmi}{\mathrm{i}}

\makeatletter
\let\MyIntOrig\int
\def\MyIntSpace{\hspace{-.25em}} 
\def\int{\MyInt}
\def\MyInt{\MyIntOrig\MyIntSkipMaybe}
\def\MyIntSkipMaybe{
  \@ifnextchar_{\MyIntSkipScript}{%
  \@ifnextchar^{\MyIntSkipScript}{%
  \@ifnextchar\limits{\MyIntSkipTok}{%
  \@ifnextchar\nolimits{\MyIntSkipTok}{%
  \MyIntSpace}}}}%
}
\def\MyIntSkipScript#1#2{#1{#2}\MyIntSkipMaybe}
\def\MyIntSkipTok#1{#1\MyIntSkipMaybe}
\makeatother

\begin{document}

\title{Twist-tuned quantum criticality in moir\'e bilayer graphene}

\author{Jan Biedermann}
\author{Lukas Janssen}

\affiliation{Institut f\"ur Theoretische Physik and W\"urzburg-Dresden Cluster of Excellence ct.qmat, TU Dresden, 01062 Dresden, Germany}

\begin{abstract}
We argue that moir\'e bilayer graphene at charge neutrality hosts a continuous semimetal-to-insulator quantum phase transition that can be accessed experimentally by tuning the twist angle between the two layers.
For small twist angles near the first magic angle, the system realizes a Kramers intervalley-coherent insulator, characterized by circulating currents and spontaneously broken time reversal and U(1) valley symmetries.
For larger twist angles above a critical value, the spectrum remains gapless down to the lowest temperatures, with a fully symmetric Dirac semimetal ground state.
Using self-consistent Hartree-Fock theory applied to a realistic model of twisted bilayer graphene, based on the Bistritzer-MacDonald Hamiltonian augmented by screened Coulomb interactions, we find that the twist-tuned quantum phase transition is continuous. 
We argue that the quantum critical behavior belongs to the relativistic Gross-Neveu-XY universality class, and we characterize it through an effective field theory analysis.
Our theoretical predictions can be directly tested using current experimental setups incorporating the recently developed quantum twisting microscope.
\end{abstract}

\date{February 25, 2025}

\maketitle


Quantum critical points denote continuous phase transitions occurring at absolute zero temperature.
They leave their imprints on finite-temperature physics through a broad quantum critical regime dominated by novel collective phenomena. 
The study of quantum critical points has been a central focus in the understanding of strongly-correlated quantum matter for decades, spanning various contexts such as unconventional superconductivity, heavy-fermion systems, and quantum magnetism~\cite{gegenwart08, sachdev11, vojta18}.

Nanoscale quantum systems composed of atomically-thin quantum materials provide an ideal platform to study quantum critical points.
Relatively simple crystal structures enable the growth of high-quality samples and facilitate well-controlled theoretical modeling.
Most importantly, nanoscale quantum systems are highly sensitive to geometric modifications, enabling the design of systems with tailored properties.
In recent years, twisted bilayer graphene has emerged as a prototype for a new class of nanoscale systems known as moir\'e materials~\cite{andrei21}. When the two graphene layers are twisted at an angle of around $1.1^\circ$, low-temperature transport experiments reveal a range of unconventional superconducting, strange metal, and correlated insulating phases as a function of electron band filling~\cite{cao18, lu19, cao20b, andrei20}.
Since the first demonstration of twisted bilayer graphene, several other moir\'e materials have been realized, including heterobilayers of transition metal dichalcogenides~\cite{tang20, regan20, wang20} and twisted multilayer systems~\cite{burg19, cao20a, shen20, park21}.
Most quantum phase transitions in these systems~\cite{jaoui22} involve electronic excitations across a range of momenta within the moir\'e Brillouin zone, a scenario that is still poorly understood theoretically~\cite{abanov03, metlitski10a, metlitski10b, liu19, lee18}.

\begin{figure}[b!]
\includegraphics[width=\linewidth]{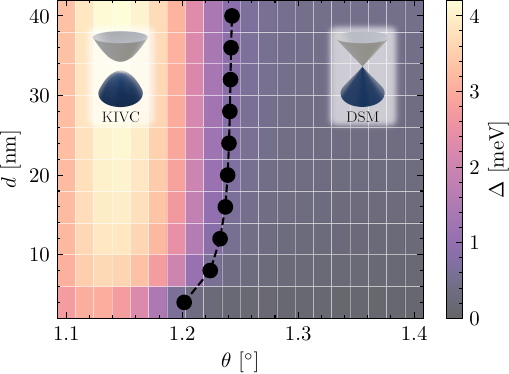}
\caption{Quantum phase diagram of moir\'e bilayer graphene as function of twist angle $\theta$ and gate distance $d$.
For small twist angles, a Kramers intervalley-coherent insulator (KIVC) is stabilized. This state breaks both time reversal and U(1) valley symmetry and features a fully gapped electronic spectrum, with the gap $\Delta$ indicated by the color scale.
Increasing the twist angle $\theta$ beyond a critical value, a transition towards a fully symmetric Dirac semimetal (DSM) is found.
The KIVC-to-DSM transition is continuous and falls into the relativistic Gross-Neveu-XY universality class.}
\label{fig:phase-diagram}
\end{figure}

In this work, we argue that moir\'e bilayer graphene at charge neutrality hosts a continuous semimetal-to-insulator quantum phase transition that allows a theoretical understanding and can be accessed experimentally by tuning the twist angle between the two layers, see Fig.~\ref{fig:phase-diagram}.
%
%
For small twist angles, we find that the system realizes a Kramers intervalley-coherent insulator (KIVC), in agreement with previous Hartree-Fock~\cite{bultinck20, kwan21}, quantum Monte Carlo~\cite{hofmann22}, and dynamical mean-field~\cite{rai24} results.
For larger twist angles above a critical value, the spectrum remains gapless down to the lowest temperatures, realizing a fully symmetric Dirac semimetal ground state.
A key simplification for theoretical understanding of this twist-tuned transition is that the electronic excitations at low energy are restricted to momenta near isolated points in the moir\'e Brillouin zone, enabling controlled field theory approaches based on the $1/N$ or $\epsilon$ expansion~\cite{gracey18, boyack21, herbut24}.
Using self-consistent Hartree-Fock theory applied to a realistic model of twisted bilayer graphene, combined with a complementary field-theory analysis, we demonstrate that the twist-tuned quantum phase transition is continuous and belongs to the relativistic Gross-Neveu-XY universality class.
In the past, quantum critical points involving Dirac fermions have been extensively studied in the theoretical literature across various contexts, including toy models for aspects of high-energy physics~\cite{braun11, gehring15, dabelow19, hands19, cresswell23}, unconventional superconductors~\cite{vojta00,huh08,schwab22}, monolayer~\cite{herbut06, herbut09b, assaad13, janssen14, otsuka16} and Bernal-stacked bilayer~\cite{pujari16, ray18, ray21b} graphene, boundary effects in topological insulators~\cite{lee07, grover14}, as well as frustrated quantum magnets~\cite{seifert20, ray21a, liu22, liu24, ray24}.
We propose that twist tuning of moir\'e materials could enable the experimental realization of this intriguing physics for the first time.

\paragraph*{Model.}
%
We employ a continuum description of moir\'e bilayer graphene at charge neutrality ($\nu = 0$) based on the Bistritzer-MacDonald model~\cite{bistritzer11}, augmented by screened Coulomb interactions. 
We project onto a subset of twelve moir\'e bands per spin species and assume empty (fully occupied) remote conduction (valence) bands~\cite{bultinck20}. In the band basis, the projected Hamiltonian reads
\begin{align} \label{eq:model}
	\mathcal{H} & = \sum_{\mathbf k \in \text{mBZ}} c^\dagger_{\mathbf k} h(\mathbf k) c_{\mathbf{k}} -
	\frac{1}{2 A} \sum_{\mathbf q} V_{\mathbf q} : \mathrel{\rho_{\mathbf q} \rho_{-\mathbf q}}:
\end{align}
where $c^\dagger_{\mathbf k} = (c^\dagger_{\mathbf k,n,\tau,s})$ creates an electronic quasiparticle characterized by band $n = 1,\dots,6$, valley $\tau = \mathbf K, \mathbf K'$, and spin $s = {\uparrow}, {\downarrow}$ indices,
and $h(\mathbf k)$ corresponds to the single-particle Hamiltonian, which contains the twist-angle-dependent Bistritzer-MacDonald dispersion for nearest-neighbor intralayer hopping $t_0 = 2.8\,\text{eV}$ and intra- and intersublattice interlayer hoppings $w_0 = 80\,\text{meV}$ and $w_1 = 110\,\text{meV}$, respectively, and a subtraction term employed to avoid double counting of interaction effects~\cite{bultinck20, liu21}.
Details are given in the appendix.
The electron density is expressed as $\rho_{\mathbf q} = \sum_{\mathbf k \in \text{mBZ}} c^\dagger_{\mathbf k} \Lambda(\mathbf k, \mathbf q) c_{\mathbf k + \mathbf q}$,
where $\Lambda_{\alpha, \beta}(\mathbf k, \mathbf q) = \langle u_{\mathbf k, \alpha} \vert u_{\mathbf k + \mathbf q, \beta} \rangle$ are the overlap matrix elements of the single-particle eigenstates $u_{\mathbf k, \alpha}$, with collective index $\alpha = (n,\tau,s)$.
In the interaction term in Eq.~\eqref{eq:model}, the colons denote normal ordering, and
$A = \frac{3\sqrt{3}}{8\sin^2{\theta}/{2}} a_0^2 L^2$ represents the total area of the sample, consisting of $L \times L$ moir\'e unit cells, with $a_0 = 0.142\,\text{nm}$ the intralayer nearest-neighbor distance and $\theta$ the twist angle.
For our parameters, the first magic angle is located at $\theta^\star = 1.068^\circ$.
We use the convention that momenta $\mathbf k$ are restricted to the first moir\'e Brillouin zone, while $\mathbf q = \mathbf G + \mathbf k$ represents unrestricted momenta in the full reciprocal space, with $\mathbf G = m_1 \mathbf G_1 + m_2 \mathbf G_2$ denoting reciprocal lattice vectors spanned by the basis vectors $\mathbf G_{1,2}$.

We adopt a screened Coulomb potential suitable for a setup with top and bottom gates, given by
\begin{align}
V_{\mathbf q} = \frac{e^2}{2\epsilon_0 \epsilon(\theta, \mathbf q) \vert \mathbf q \vert} \tanh(\vert \mathbf q\vert d),
\end{align}
where $e$ is the electron charge and $\epsilon(\theta, \mathbf q)$ is an effective permittivity that accounts for both the dielectric screening of the substrate, assumed to be $\epsilon_\text{substrate} = 5$, and the twist-angle-dependent internal screening from graphene electrons~\cite{goodwin19, goodwin20a, goodwin20b}. The explicit form of $\epsilon(\theta, \mathbf q)$ is given in the appendix.
$d$ is the distance between the sample and the gates, which is assumed to be $d = 20\,\text{nm}$ unless stated otherwise.
%
%
To simplify the numerics, the potential $V_{\mathbf q}$ is truncated for large momenta $\mathbf q = \mathbf G + \mathbf k$ with $|\mathbf G| > 2 |\mathbf G_{1,2}|$.

The effective model defined in Eq.~\eqref{eq:model} features an emergent $\mathrm{U}(2)_{\mathbf K} \times \mathrm{U}(2)_{\mathbf K'} \simeq \mathrm{U}(1)_{\text{charge}} \times \mathrm{U}(1)_{\text{valley}} \times \mathrm{SU}(2)_{\mathbf K} \times \mathrm{SU}(2)_{\mathbf K'}$ symmetry.
Here, $\mathrm{U}(1)_{\text{charge}}$ and $\mathrm{U}(1)_{\text{valley}}$ represent charge and valley charge conservation, generated by $\mathbb{1}$ and $\tau_z$, respectively, with the Pauli matrix $\tau_z$ acting in valley space, and
$\mathrm{SU}(2)_{\mathbf{K}}$ and $\mathrm{SU}(2)_{\mathbf{K}'}$ correspond to independent spin rotations in the two valleys.
At the lattice scale, these symmetries are present only in the limit of small twist angles. For finite twist angles, $\mathrm{SU}(2)_{\mathbf K} \times \mathrm{SU}(2)_{\mathbf K'}$ reduces to global $\mathrm{SU}(2)$ spin rotations and $\mathrm{U}(1)_{\text{valley}}$ breaks down to $\mathbb Z_3$~\cite{wu19, bultinck20}.

\paragraph*{Electronic spectrum.}

\begin{figure}[tb!]
\includegraphics[width=\linewidth]{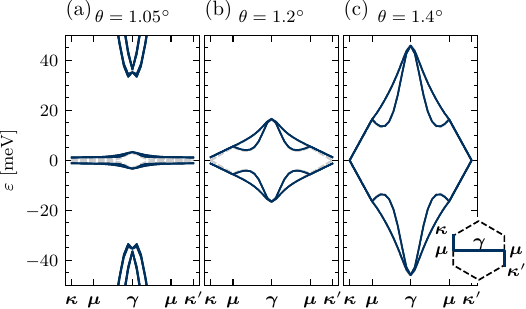}
\caption{%
(a)~Electronic spectrum along a high-symmetry path in the moir\'e Brillouin zone for twist angle $\theta = 1.05^\circ$ in the KIVC phase near the first magic angle, from Hartree-Fock theory for $L = 18$ (solid blue lines). At charge neutrality, the Fermi level is located at $\varepsilon = 0$. The spectrum is fully gapped, with the minimal gap occurring at the corners $\boldsymbol{\kappa}$ and $\boldsymbol{\kappa}'$ of the moir\'e Brillouin zone.
For comparison, the corresponding noninteracting bands are also shown (dashed gray lines).
(b)~Same as~(a), but for twist angle $\theta = 1.2^\circ$ still within the KIVC phase, but closer to the transition.
(c)~Same as~(a), but for twist angle $\theta = 1.4^\circ$ in the Dirac semimetal phase. The spectrum is gapless, with four bands per spin species crossing at the $\boldsymbol\kappa$ and $\boldsymbol\kappa'$ points. The inset indicates the high-symmetry path in the moiré Brillouin zone along which the dispersions are plotted.
}
\label{fig:spectrum}
\end{figure}

\begin{figure*}[tb!]
\includegraphics[width=\linewidth]{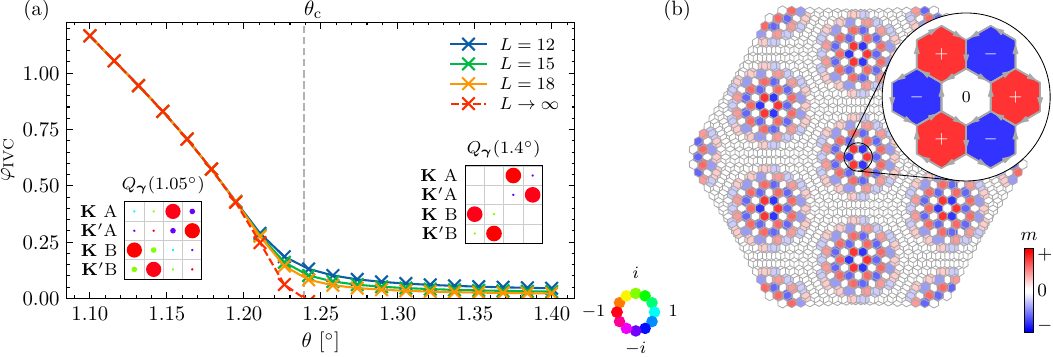}
\caption{%
(a)~Order parameter for intervalley coherence $\varphi_\text{IVC}$ as a function of twist angle $\theta$ for different system sizes~$L$. For $\theta$ approaching $\theta_\mathrm{c}$ from below, the order parameter vanishes continuously. The extrapolation towards the thermodynamic limit (dashed red line) suggests a quantum critical point at $\theta_\mathrm{c} \approx 1.24^\circ$, in agreement with the value obtained from the crossing point analysis presented in Fig.~\ref{fig:crossing} (gray dashed line).
The two different ground states are characterized by different single-electron density matrices $P (\mathbf k) = [\mathbb{1} + Q(\mathbf k)]/2$, with $Q_{\boldsymbol{\gamma}} \equiv Q(\mathbf k = \mathbf 0)$ shown in the valley-sublattice basis $(\tau = \mathbf K, \mathbf K', \sigma = \mathrm{A}, \mathrm{B})$ for $\theta = 1.05^\circ$ (KIVC insulator) and $\theta = 1.4^\circ$ (Dirac semimetal) in the insets. Here, the radii (colors) of the dots indicate the magnitudes (phases) of the corresponding matrix entries.
(b)~Schematic representation of the KIVC state across multiple real-space moir\'e unit cells. Arrows indicate the alternating circulating currents for a single spin species. The colored plaquettes show the associated magnetic flux pattern, which triples the monolayer unit cell, visible in the close-up view shown in the inset. The magnitude of the magnetization density $m$ is maximal (minimal) at AA (AB) regions in the moir\'e unit cell.
}
\label{fig:kivc}
\end{figure*}

We use self-consistent Hartree-Fock theory~\cite{guinea18, cea20, bultinck20, kwan21, liu21, cea22} to compute the ground state of the Hamiltonian in Eq.~\eqref{eq:model} at charge neutrality for various twist angles $\theta$.
Technical details of the method are provided in the appendix.
Figure~\ref{fig:spectrum} shows the electronic spectrum (solid blue lines) along a high-symmetry path in the moir\'e Brillouin zone for different twist angles $\theta$. For comparison, the corresponding bands in the noninteracting limit are also shown (dashed gray lines).
At a small twist angle of $\theta = 1.05^\circ$, near the first magic angle [Fig.~\ref{fig:spectrum}(a)], the noninteracting spectrum features four flat bands per spin species. Including interactions, the flat bands split into two valence bands at negative energy and two conduction bands at positive energy, separated by a full spectral gap. 
The minimal gap occurs at the corners $\boldsymbol\kappa$ and $\boldsymbol\kappa'$ of the moir\'e Brillouin zone, which correspond to the crossing points of the low-energy bands in the noninteracting limit.
For $\theta = 1.05^\circ$, we find a gap of $\Delta \simeq 2.31\,\text{meV}$, which is of the same order of magnitude as the experimentally observed value $\Delta \simeq 0.86\,\text{meV}$ in magic-angle twisted bilayer graphene~\cite{lu19}. Note that previous Hartree-Fock studies, which neglected internal screening effects, consistently overestimated the interaction-induced gap size, often substantially~\cite{cea20, bultinck20, kwan21, hofmann22, cea22}.
At a slightly larger twist angle of $\theta = 1.2^\circ$ [Fig.~\ref{fig:spectrum}(b)], the width of the noninteracting bands at low energy increases, but the interacting electronic spectrum remains fully gapped.
At an even larger twist angle of $\theta = 1.4^\circ$ [Fig.~\ref{fig:spectrum}(c)], the interaction-induced band renormalization is largely suppressed, causing the interacting and noninteracting spectra to nearly coincide.
In particular, the interacting spectrum is now gapless (up to finite-size corrections), and the system realizes a Dirac semimetal ground state, with four bands per spin species crossing at the corners $\boldsymbol\kappa$ and $\boldsymbol\kappa'$ of the moir\'e Brillouin zone.

The size of the spectral gap $\Delta$ and the critical value of $\theta$, at which the gap closes, depend on the gate distance~$d$. However, we have confirmed that the same qualitative behavior persists across a range of $d$ values from $4\,\text{nm}$ to $40\,\text{nm}$.
The spectral gap $\Delta$ as a function of $\theta$ and $d$ is depicted using a color scale in the quantum phase diagram shown in Fig.~\ref{fig:phase-diagram}.

\paragraph*{Ground state.}

To characterize the ground state as a function of twist angle, we analyze the density matrix $P_{n,\tau,s;n',\tau',s'}(\mathbf k) = \langle c^\dagger_{\mathbf k, n',\tau',s'} c_{\mathbf k, n,\tau,s} \rangle$. This enables the construction of the order parameter for intervalley coherence as
$\varphi_\text{IVC} = L^{-2}\sum_{\mathbf k} \sqrt{\operatorname{Tr} [P_{\text{IVC}}(\mathbf k)^2 ]}$,
where $P_\text{IVC}$ is the $\mathrm{U}(1)_\text{valley}$ breaking part of the density matrix, defined as $P_{\text{IVC } n, \tau, s; n', \tau', s'}(\mathbf k) =  (1 - \delta_{\tau \tau^\prime}) P_{n, \tau, s; n', \tau', s'}(\mathbf k)$~\cite{bultinck20}.
Figure~\ref{fig:kivc}(a) shows $\varphi_\text{IVC}$ as a function of twist angle $\theta$ for different system sizes $L$. For each value of $\theta$, we perform a linear extrapolation of $\varphi_\text{IVC}$ as a function of $1/L$ towards the thermodynamic limit $1/L \to 0$, resulting in the red crosses shown in Fig.~\ref{fig:kivc}(a).
The behavior of $\varphi_\text{IVC}$ as a function of $\theta$ in the thermodynamic limit demonstrates that the $\mathrm{U}(1)_\text{valley}$ symmetry is spontaneously broken for $\theta < \theta_\mathrm{c}$ and remains intact for $\theta > \theta_\mathrm{c}$, with $\theta_\mathrm{c} \approx 1.24^\circ$.

For twelve spin-degenerate bands in the Hartree-Fock calculation, the density matrix $P(\mathbf k)$ is a $12 \times 12$ matrix for each spin species.
To determine the nature of the symmetry-broken state, we examine the off-diagonal part $Q(\mathbf k)$ of the density matrix, defined implicitly as $P(\mathbf k) = [\mathbb{1} + Q(\mathbf k)]/2$, within the subspace of the four low-energy bands.
For interpretation, it is convenient to rotate the basis such that the sublattice operator $\sigma_{nn'} = \langle u_n(\mathbf k)|\sigma_z |u_{n'}(\mathbf k) \rangle$, where $\sigma_z$ is the Pauli matrix acting in sublattice space, becomes diagonal.

For a fully symmetric Dirac semimetal ground state, the $4\times 4$ matrix $(Q_{\tau,\sigma;\tau',\sigma'})$ is of the form~\cite{liu21}
\begin{align} \label{eq:Q_DSM}
Q_\text{DSM}(\mathbf k) = \sigma_x \mathrm{e}^{\mathrm{i}\alpha_{\mathbf k} \sigma_z \tau_z}
\stackrel{\mathbf k = \mathbf 0}{\longrightarrow}
\raisebox{-.58cm}{\,\includegraphics{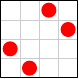}\,}\,,
\end{align}
where the Pauli matrices $\vec \sigma$ and $\vec \tau$ act in sublattice space and valley space, respectively. 
The phase $\alpha_{\mathbf k}$ winds around the gapless points by $\pm 2\pi$.
In the graphical representation of $Q_\text{DSM}$, we have set $\alpha_{\mathbf k = \mathbf 0} = \pi$ for ease of comparison with the numerical result discussed below, with the color code provided in the inset of Fig.~\ref{fig:kivc}(a).

In the strong-coupling limit, the KIVC state is characterized by the $4\times 4$ matrix
\begin{align}
\label{eq:Q_KIVC}
Q_\text{KIVC} & = \sigma_y (\tau_x \cos \vartheta + \tau_y \sin \vartheta) 
\stackrel{\vartheta = 0}{\longrightarrow}
\raisebox{-.58cm}{\,\includegraphics[scale=1]{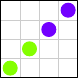}\,}\,,
\end{align}
where we have set $\vartheta = 0$ in the graphical representation.
Away from the strong-coupling limit, the density matrix will receive finite semimetallic contributions parallel to $Q_\text{DSM}$, arising from the band dispersion.
The KIVC order breaks time reversal $\mathcal T = \tau_x \mathcal K$ and $\mathrm{U}(1)_\text{valley}$ symmetry, but preserves a Kramers time reversal defined as $\mathcal T' = \tau_y \mathcal K$, with $\mathcal K$ corresponding to complex conjugation.
In real space, the KIVC order features alternating circulating currents around the elemental plaquettes of the graphene unit cell, with the currents reaching their maximum (minimum) at the AA (AB) regions within the moir\'e unit cell~\cite{bultinck20}.
The resulting magnetization pattern is schematically illustrated in Fig.~\ref{fig:kivc}(b).

The insets of Fig.~\ref{fig:kivc}(a) show the off-diagonal parts $Q_{\boldsymbol \gamma} \equiv Q(\mathbf k = \mathbf 0)$ of the density matrices obtained from self-consistent Hartree-Fock calculations restricted to the four low-energy bands per spin species, for two different values of the twist angle $\theta$.
At $\theta = 1.05^\circ$, $Q_{\boldsymbol \gamma}$ is consistent with the KIVC form described in Eq.~\eqref{eq:Q_KIVC}, demonstrating that the gap observed in the electronic spectrum for $\theta < \theta_\mathrm{c}$ originates from a KIVC ground state.
This result agrees with previous Hartree-Fock~\cite{bultinck20, kwan21}, quantum Monte Carlo~\cite{hofmann22}, and dynamical mean-field~\cite{rai24} calculations.
At $\theta = 1.4^\circ$, however, the $\mathrm U(1)_\text{valley}$-symmetry-breaking matrix entries of $Q_{\boldsymbol \gamma}$ are negligibly small and further diminish with increasing system size $L$, consistent with the Dirac semimetal form described in Eq.~\eqref{eq:Q_DSM}.
This confirms that the ground state for $\theta > \theta_\mathrm{c}$ is a fully symmetric Dirac semimetal, in agreement with our conclusion based on the behavior of the order parameter $\varphi_\text{IVC}$.

\paragraph*{Twist-tuned transition.}

\begin{figure}[tb!]
\includegraphics[width=\linewidth]{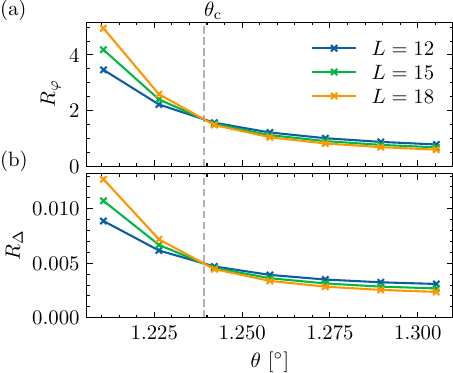}
\caption{%
(a)~Crossing-point analysis of the order-parameter renormalization group invariant $R_\varphi = L \varphi_\text{IVC}$ as function of twist angle~$\theta$ for different system sizes $L$. The crossing at $\theta_\mathrm{c} = 1.24(1)^\circ$
(dashed gray line) indicates the location the quantum critical point between the KIVC insulator for $\theta < \theta_\mathrm{c}$ and the Dirac semimetal for $\theta > \theta_\mathrm{c}$.
(b)~Same as (a), but for the gap renormalization group invariant $R_\Delta = L \Delta/t_0$, which reveals a consistent crossing point location.}
\label{fig:crossing}
\end{figure}

As shown in Fig.~\ref{fig:kivc}(a), the order parameter $\varphi_\text{IVC}$ vanishes continuously as $\theta$ approaches $\theta_\mathrm{c}$ from below.
This suggests a continuous twist-tuned quantum phase transition between a symmetry-broken KIVC insulator and a symmetric Dirac semimetal.
This scenario can be corroborated through a finite-size scaling analysis~\cite{campostrini14}.
To this end, we construct dimensionless renormalization group invariants $R_\varphi = L^{(1+\eta_\varphi)/2} \varphi_\text{IVC}$ and $R_\Delta = L^{z} \Delta/t_0$ from the order parameter $\varphi_\text{IVC}$ and the electronic spectral gap $\Delta$, with exponents $z = 1$ and $\eta_\varphi = 1$ at the Hartree-Fock level.
For a continuous twist-tuned transition, the curves for $R_\varphi$ and $R_\Delta$ as functions of $\theta$ for different fixed system sizes $L$ will cross at the critical twist angle $\theta_\mathrm{c}$, provided that corrections to scaling can be neglected.
Figure~\ref{fig:crossing} shows the crossing-point analysis for $d = 20\,\text{nm}$. The curves clearly exhibit a single crossing at $\theta_\mathrm{c} = 1.24(1)^\circ$, further supporting the continuous nature of the transition.

\paragraph*{Quantum critical behavior.}

The existence of a single divergent length scale near the quantum critical point enables us to describe the low-energy behavior by expanding around momenta near the $\boldsymbol{\kappa}$ and $\boldsymbol{\kappa}'$ points in the moir\'e Brillouin zone, where the noninteracting bands cross~\cite{christos21, parthenios23}.
This way, we obtain a low-energy field theory for the 16-component Dirac spinor $\psi_{\mathbf k} = (c_{n,\tau,s}(\boldsymbol{\kappa} + \mathbf k), c_{n,\tau,s}(\boldsymbol{\kappa}' + \mathbf k))$ and the two-component real order-parameter field $\vec{\varphi} = (\varphi^1, \varphi^2)$, with Euclidean action
\begin{align} \label{eq:field-theory}
	S = \int \dd[2]{\mathbf x} \dd{\tau} \left( \bar\psi \gamma_\mu \partial_\mu \psi + \varphi_1 \rmi \bar\psi \gamma_3 \psi + \varphi_2 \rmi \bar\psi \gamma_5 \psi \right).
\end{align}
Here, $\gamma_\mu$ with $\mu = 0,1,2$ denote $16\times 16$ Dirac matrices satisfying the Clifford algebra $\{\gamma_\mu, \gamma_\nu\} = 2 \delta_{\mu\nu}$, and we have assumed summation convention over repeated space-time indices.
The matrices $\gamma_3$ and $\gamma_5$ square to one and anticommute with $\gamma_\mu$ and with each other.
They are chosen such that the Hermitian product $\gamma_{35} = \rmi \gamma_3 \gamma_5$ generates $\mathrm{U}(1)_\text{valley}$ transformations in spinor space.
As a consequence, the order parameter $\vec \varphi$ transforms as a vector under $\mathrm{U}(1)_\text{valley}$ transformations, and a nonvanishing expectation value $\varphi_\text{IVC} \propto |\langle \vec \varphi \rangle|$ signals intervalley coherence and spontaneous $\mathrm{U}(1)_\text{valley}$ symmetry breaking.
At finite twist angles, the continuous $\mathrm{U}(1)_\text{valley}$ symmetry is only approximately realized at the lattice scale. At the quantum critical point, however, the symmetry breaking terms have been argued to be irrelevant in the renormalization group sense, such that the $\mathrm{U}(1)_\text{valley}$ symmetry emerges at low energy~\cite{li17, classen17}.

The quantum field theory given in Eq.~\eqref{eq:field-theory} features a critical point that falls into the Gross-Neveu-XY universality class~\cite{roy10, zerf17, janssen18}. 
Due to the presence of gapless fermions at the transition, this universality class extends beyond the Landau-Ginzburg-Wilson paradigm, which is based solely on an order-parameter description~\cite{vojta18}.
The universal behavior that characterizes the fermionic quantum criticality can be accessed within the $1/N$ expansion, where $N$ corresponds to the number of four-component Dirac spinors~\cite{gracey18, gracey21}.
For the order-parameter anomalous dimension $\eta_\varphi$, governing the form of the dynamical structure factor $\mathcal S(\mathbf k, \omega) \propto 1/(\omega^2 - \mathbf k^2)^{(2-\eta_\varphi)/2}$ at the critical point $\theta = \theta_\mathrm{c}$, we find
\begin{align}
\eta_\varphi = 1 - \frac{8}{3\pi^2N} + \frac{544}{27\pi^4 N^2} + \mathcal O(1/N^3) \simeq 0.945,
\end{align}
where we have set $N = 4$, relevant for the present case. Notably, $\eta_\varphi$ is more than an order of magnitude larger than its counterpart in the conventional 2+1D XY universality class~\cite{chester20}, highlighting the significant deviation from the Landau-Ginzburg-Wilson paradigm.
For the exponent $\nu$, governing the divergence of the correlation length $\xi \propto |\theta - \theta_\mathrm{c}|^{-\nu}$, we obtain~\footnote{%
Note that Eq.~(8.6) in Ref.~\cite{gracey21} contains a typo in the $\mathcal O(1/N^2)$ contribution to $1/\nu$, where it should read $\pi^4$ instead of $\pi^6$.}
\begin{align}
1/\nu = 1 - \frac{16}{3\pi^2 N} + \frac{216\pi^2 + 2912}{27\pi^4 N^2} + \mathcal O(1/N^3) \simeq 0.985,
\end{align}
The fermion anomalous dimension $\eta_\psi$, governing the form of the fermion spectral function $A(\mathbf k, \omega) \propto 1/(\omega^2 - \mathbf k^2)^{(1-\eta_\psi)/2}$ at the critical point, is given by
\begin{align}
\eta_\psi = \frac{4}{3\pi^2 N} + \frac{160}{27\pi^4 N^2} - \frac{80(3\pi^2+20)}{243\pi^6N^3} + \mathcal O(1/N^4) \simeq 0.037.
\end{align}
The power-law behavior of $A(\mathbf k, \omega)$ implies a continuum of fermion excitations without a well-defined quasiparticle description.
This signifies the existence of a quantum critical regime at finite temperatures characterized by non-Fermi liquid behavior with nontrivial exponents~\cite{loehneysen07}. On both sides of the transition, the quantum critical regime is separated from the respective low-temperature phase by a crossover at $T^\star \propto |\theta - \theta_\mathrm{c}|^{\nu z}$, where $z$ is the dynamical critical exponent. Owing to the emergent relativistic symmetry at the quantum critical point~\cite{herbut09a, roy16}, we obtain $z=1$.
The remaining critical exponents follow from hyperscaling relations~\cite{herbutbook}. In particular, we find that the order parameter for $\theta < \theta_\mathrm{c}$ scales as $\varphi_\text{IVC} \propto (\theta_\mathrm{c} - \theta)^\beta$ with $\beta \simeq 0.988$, close to the Hartree-Fock mean-field value $\beta = 1$, observed in Fig.~\ref{fig:kivc}(a).

For $N=4$, the large-$N$ corrections to the exponents are small, indicating that order-parameter fluctuations are subdominant. This implies that the Hartree-Fock analysis, which is controlled in the limit $N \to \infty$, yields reasonably accurate results even for the physical case of $N = 4$.

\paragraph*{Conclusions.}

We have argued that moir\'e bilayer graphene at charge neutrality hosts a continuous semimetal-to-insulator quantum phase transition that falls into the relativistic Gross-Neveu-XY universality class.
The transition can be experimentally probed by tuning the twist angle between the two layers. 
In particular, we propose that a setup employing the recently developed quantum twisting microscope~\cite{inbar23} could provide a means to verify our theoretical predictions, potentially achieving the first experimental realization of a relativistic fermionic quantum critical point.
Alternatively, the transition can be driven by adjusting the gate distance for a fixed twist angle. Another possibility is applying hydrostatic pressure~\cite{yankowitz18,yankowitz19}, which we expect to increase the critical angle $\theta_\mathrm{c}$ by enhancing internal screening through the reduction of the low-energy band width.
Similar relativistic Gross-Neveu-type quantum phase transitions might also be realized in other moir\'e materials, a possibility that warrants further investigation in future studies.
 
\begin{acknowledgments}
\paragraph{Acknowledgments.}
We thank
Igor Herbut,
Johannes Hofmann,
Zi Yang Meng,
Michael Scherer,
and
Tim Wehling
for valuable discussions.
This work has been supported by the Deutsche Forschungsgemeinschaft (DFG) through 
Project No.\ 247310070 (SFB 1143, A07), 
Project No.\ 390858490 (W\"urzburg-Dresden Cluster of Excellence \textit{ct.qmat}, EXC 2147), and 
Project No.\ 411750675 (Emmy Noether program, JA2306/4-1).

\paragraph{Data availability.}

The data that support the findings of this article are openly available~\cite{data-availability}.

\paragraph*{Note added.}
During the preparation of this manuscript, two related preprints appeared on arXiv.
Ref.~\cite{ma24} reports the experimental observation of a twist-tuned semimetal-to-insulator transition in moir\'e WSe$_2$ tetralayers upon hole doping. We anticipate that a suitably adapted version of our theory also accounts for the experimental results presented in that study.
Ref.~\cite{huang24} reports a quantum Monte Carlo study of the twist-tuned transition in moir\'e bilayer graphene. The results are consistent with ours, accounting for differences in the form of the screened Coulomb potential (single- vs.\ double-gated setup), the values of the hopping parameters, and the number of bands taken into account (four vs.\ twelve bands per spin species).

\end{acknowledgments}

\FloatBarrier
\bibliographystyle{longapsrev4-2}
\bibliography{tbg-crit}


\title{\vspace{\paperheight} 
End Matter}

\makeatletter
\def\@date{} 
\patchcmd{\frontmatter@RRAP@format}{(}{}{}{} 
\patchcmd{\frontmatter@RRAP@format}{)}{}{}{}
\title@column\titleblock@produce 
\makeatother

\setcounter{equation}{0}
\renewcommand\theequation{A\arabic{equation}}


\paragraph{Bistritzer-MacDonald Hamiltonian.}

In this section, we review the Bistritzer-MacDonald model~\cite{bistritzer11} within the different bases employed.
The Hamiltonian for twisted bilayer graphene may be expressed in the continuum limit as
\begin{align} \label{eq:BM-model}
\mathcal{H}_{\text{BM}} &= \sum_{\mathbf p} f^\dagger_{\mathbf p} h_\mathrm{D}(\mathbf p) f_{\mathbf p} + \sum_{\mathbf p, \mathbf p^\prime} \sum_{j = 0}^2 f^\dagger_{\mathbf p^\prime} \mu_x T_j \delta_{\mathbf p^\prime, \mathbf p - \tau_z \mathbf q_j} f_{\mathbf p}\,,
\end{align}
where the sums over $\mathbf p$ and $\mathbf p'$ extend over all momenta in reciprocal space,
and $f_{\mathbf p} = (f_{\mathbf p, \ell, \sigma, \tau, s})$ annihilates an electron characterized by layer $\ell = \mathrm{t}, \mathrm{b}$, sublattice $\sigma = \mathrm{A}, \mathrm{B}$, valley $\tau = \mathbf K, \mathbf K'$, and spin $s = {\uparrow}, {\downarrow}$ indices.

\begin{figure}[b!]
\includegraphics{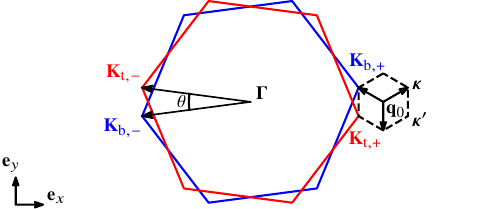}
\caption{Schematic representation of the single-layer Brillouin zones of top (red) and bottom (blue) layers, along with the resulting moir\'e Brillouin zone (black).}
\label{fig:brillouin-zone}
\end{figure}

The first term in Eq.~\eqref{eq:BM-model} represents the continuum Dirac Hamiltonian for the individual, decoupled layers, with
\begin{align}
[h_\mathrm{D}(\mathbf p)]_{\ell,\tau;\ell',\tau'} = \hbar v_\text{F} (\mathbf p - \mathbf K_{\ell,\tau}) \cdot (\tau \sigma_x, \sigma_y)\delta_{\ell \ell^\prime} \delta_{\tau \tau^\prime} \,,
\end{align}
where the Pauli matrices $\sigma_{x,y}$ act in sublattice space.
The Fermi velocity of the single layer is given by $v_\text{F} = 3 a_0 t_0/(2\hbar)$, where $a_0 = 0.142\,\text{nm}$ is the intralayer nearest-neighbor distance and $t_0 = 2.8\,\text{eV}$ is the corresponding hopping parameter.
$\mathbf K_{\ell, \tau}$ corresponds to the Dirac point of the layer $\ell$ and valley $\tau$, given as
\begin{align}
\mathbf K_{\ell,\tau} = \tau R_{-\ell \theta / 2} \mathbf K = \tau R_{-\ell \theta / 2} \frac{4 \pi}{3 \sqrt{3} a_0} \mathbf e_x\,.
\end{align}
In the above, with a slight abuse of notation, we take $\ell = +1$ ($\ell = -1$) for the top (bottom) layer and $\tau = + 1$ ($\tau = -1$) for the $\mathbf K$ ($\mathbf K'$) valley.
The location of the Dirac points $\mathbf K_{\ell, \tau}$, along with the single-layer Brillouin zone and the resulting moir\'e Brillouin zone, is illustrated in Fig.~\ref{fig:brillouin-zone}.

The second term in Eq.~\eqref{eq:BM-model} represents the interlayer hopping, with Hermitian hopping matrices
\begin{align}
T_j &= w_0 \sigma_0 + w_1 \left( \sigma_x \cos \tfrac{2 \pi j}{3} + \tau_z  \sigma_y \sin \tfrac{2 \pi j}{3} \right),
\end{align}
$j = 0,1,2$, with intra- and intersublattice interlayer hopping parameters $w_0 = 80\,\text{meV}$ and $w_1 = 110\,\text{meV}$~\cite{bultinck20}.
The Pauli matrix $\mu_x$ in Eq.~\eqref{eq:BM-model} acts in layer space,
while $\tau_z$ acts in sublattice space.
The momentum transfer for the interlayer hopping is given by $\mathbf q_j = R_{2\pi j/3} \mathbf q_0$, where $R_\alpha$ denotes a rotation by an angle $\alpha$, and $\mathbf q_0 = K_{\text{t},+} - K_{\text{b},+}$. The momentum transfer vectors $\mathbf q_{0,1,2}$ are indicated by black arrows in Fig.~\ref{fig:brillouin-zone}.

For the numerical analysis, it is convenient to rotate the basis of the Hilbert space so that the band index becomes a good quantum number.
This is achieved by introducing electron creation operators as
\begin{align}
c^\dagger_{\mathbf k, n, \tau, s} = \sum_{\mathbf G, \sigma, \ell} u_{n, \tau; \mathbf G, \sigma, \ell}(\mathbf k) f^\dagger_{\mathbf k + \mathbf G, \sigma, \ell, \tau, s}
\end{align}
and associated annihilation operators $c_{\mathbf k, n, \tau, s}$,
where the rotation matrix elements $u_{n, \tau; \mathbf G, \sigma, \ell}(\mathbf k)$ are obtained by diagonalizing the Hamiltonian in a truncated plane-wave basis. Specifically, we expand the wavefunctions as
\begin{align}
\vert u_{\mathbf k, n, \tau} (\mathbf r) \rangle = \sum_{\substack{\mathbf G, \sigma, \ell \\ |\mathbf G| \leq 3 \vert \mathbf G_{1, 2} \vert}} a_{\mathbf k, n, \tau; \mathbf G, \sigma, \ell} \rme^{\rmi (\mathbf k + \mathbf K_{\ell,\tau} + \mathbf G) \cdot \mathbf r} \vert \sigma, \ell \rangle.
\end{align}
The cutoff $|\mathbf G| \leq 3 \vert \mathbf G_{1, 2} \vert$ leads to 148 noninteracting bands per spin and valley degree of freedom.

To obtain the density matrices shown in Eqs.~\eqref{eq:Q_DSM} and \eqref{eq:Q_KIVC} in the sublattice basis, we have fixed the time reversal operator as $\mathcal T = \tau_x \mathcal K$ and the particle-hole transformation as $\mathcal P = \tau_z \sigma_y \mathcal K$, where $\mathcal K$ corresponds to complex conjugation.

\paragraph{Effective permittivity.}

\begin{figure}[t]
\includegraphics{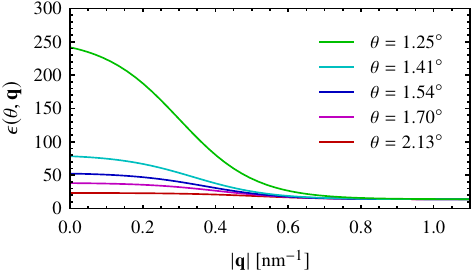}
\caption{Effective permittivity $\epsilon(\theta,\mathbf q)$ as a function of $|\mathbf q|$ for different twist angles $\theta$ above the first magic angle $\theta^\star$ from phenomenological formula given in Eq.~\eqref{eq:screening-phenomenological}, using the parameters $c_1 \simeq 0.305$, $c_2 \simeq 3.38$, and $c_3 \simeq 39.0$. The curves closely match the numerical results from the random-phase approximation of an atomistic model, with a deviation of typically less than 5\%, cf.~Fig.~2(c) of Ref.~\cite{goodwin19}.}
\label{fig:screening}
\end{figure}

In this section, we provide details of the form of the screened Coulomb interaction.
We model the static dielectric function relevant for the electrons in the twisted bilayer graphene sample within a random-phase approximation as~\cite{goodwin19}
\begin{align}
\epsilon(\theta,\mathbf q) = \epsilon_\text{substrate} + \frac{e^2}{2 \epsilon_0 |\mathbf q|} \Pi_0(\mathbf q),
\end{align}
where $\epsilon_\text{substrate}$ corresponds to the relative permittivity of the substrate and $\Pi_0(\mathbf q)$ denotes the independent-particle polarizability of the graphene electrons.
Throughout this work, we assume $\epsilon_\text{substrate} = 5$, which corresponds to the effective permittivity of hexagonal boron nitride in the static limit~\cite{tuan18, laturia18, segura18, pierret22}.
The independent-particle polarizability of the graphene electrons can be computed within an atomistic tight-binding model~\cite{goodwin19, goodwin20a, goodwin20b}. 
To simplify the calculation, we adopt the phenomenological formula from Ref.~\cite{goodwin19},
\begin{align} \label{eq:screening-phenomenological}
\epsilon(\theta,\mathbf q) = \epsilon_\text{substrate} + \epsilon_\text{decoupled} + \frac{\epsilon_\text{moir\'e}(\theta) - \epsilon_\text{decoupled}}{1 + \rme^{\left(|\mathbf q| - q_0(\theta)\right) \ell}}\,.
\end{align}
Here, $\epsilon_\text{decoupled} = 1 + e^2/(6 \epsilon_0 t_0 a_0)$ describes the permittivity of decoupled graphene bilayers~\cite{ando06}, while $\epsilon_\text{moir\'e}(\theta) = c_1/|\theta - \theta^\star|$ represents screening from moir\'e bands in the long-wavelength limit, with $\theta^\star$ denoting the first magic angle.
The crossover momentum scale $q_0(\theta) = c_2/a_\text{M}(\theta)$ is determined by the moir\'e unit cell size, $a_\text{M}(\theta) = \sqrt{3} a_0 / (2 \sin \frac{\theta}{2})$, with the associated decay length $\ell = c_3 \sqrt{3} a_0$.
The phenomenological model contains three dimensionless fitting parameters $c_1$, $c_2$, and $c_3$. ($c_1$ is dimensionless when twist angles are measured in radians.)
The numerical results from the atomistic model in Ref.~\cite{goodwin19} can be accurately fitted across a range of twist angles above the magic angle and for all momenta $\mathbf{q}$ using a fixed set of fitting parameters: $c_1 \simeq 0.305$, $c_2 \simeq 3.38$, and $c_3 \simeq 39.0$.
Figure~\ref{fig:screening} displays the resulting effective permittivity for various fixed twist angles. To facilitate comparison with Ref.~\cite{goodwin19}, here we set $\theta^\star = 1.18^\circ$, which corresponds to the first magic angle obtained using the parameters from that work. For the results presented in the main text, we have used $\theta^\star = 1.068^\circ$, corresponding to the first magic angle determined for our parameters.

While accounting for internal screening effects is essential for accurately determining the interaction-induced band gap $\Delta$ and the critical angle $\theta_\mathrm{c}$, we have explicitly verified that the qualitative features of the phase diagram and the nature of the twist-tuned transition remain unchanged when neglecting internal screening and assuming a constant permittivity $\epsilon$ over a range of values.

\paragraph{Hartree-Fock analysis.}

In this section, we provide technical details of the Hartree-Fock analysis.
To facilitate the numerical calculation, we project the Hamiltonian into a number of $N_\text{bands}$ of active bands per spin and valley degree of freedom.
We typically set $N_\text{bands} = 6$, with the exception of the density matrices displayed in the insets of Fig.~\ref{fig:kivc}(a), where $N_\text{bands} = 2$ is used.
In order to avoid double counting of interaction effects~\cite{bultinck20, liu21}, we add a subtraction term, determined by a reference density matrix $P_0$, leading to the effective Hamiltonian
\begin{align}
\mathcal{H_{\text{eff}}}[P](\mathbf k) = h_{\text{BM}}(\mathbf k) + h_{\text{H}}[P - P_0](\mathbf k) + h_{\text{F}}[P - P_0](\mathbf k),
\end{align}
where $h_{\text{BM}}$ corresponds to the Bistritzer-MacDonald Hamiltonian in the band basis, and the Hartree and Fock contributions are given as
\begin{align}
h_{\text{H}}[P](\mathbf k) &= \frac{1}{A} \sum_{\mathbf G} V_{\mathbf G} \Lambda(\mathbf k, \mathbf G) \sum_{\mathbf k^\prime} \operatorname{Tr}\left( P(\mathbf k^\prime) \Lambda(\mathbf k^\prime, \mathbf G)^\dagger \right)
\end{align}
and
\begin{align}
h_{\text{F}}[P](\mathbf k) &= - \frac{1}{A} \sum_{\mathbf q} V_{\mathbf q} \Lambda(\mathbf k, \mathbf q) P(\mathbf k + \mathbf q) \Lambda(\mathbf k, \mathbf q)^\dagger,
\end{align}
where $A = \frac{3\sqrt{3}}{8\sin^2{\theta}/{2}} a_0^2 L^2$ represents the total area of the sample, consisting of $L \times L$ moir\'e unit cells.
We take the reference density matrix $P_0$ to be the density matrix of the decoupled bilayer for $w_0 = w_1 = 0$ at charge neutrality.
Assuming that the remote band contributions of $P$ and $P_0$ cancel, it then suffices to consider the effective Hamiltonian in the subspace of the active bands.
The Hartree-Fock energy is given as
\begin{align}
E_{\text{HF}} &= \sum_{\mathbf k} \operatorname{Tr}\Bigl\{ P \Bigl(
h_{\text{BM}}(\mathbf k) 
+ h_{\text{H}}[\tfrac12 P - P_0](\mathbf k)
\nonumber \\ & \quad
+ h_{\text{F}}[\tfrac12 P - P_0](\mathbf k) \Bigr) \Bigr\}\,,
\end{align}
where the factors of $1/2$ prevent double counting of the interaction energy.

\end{document}